\documentclass[twocolumn]{pasj00}

\Received{2013 May 27}
\Accepted{2013 August 13}
\SetRunningHead{N. Sakai et al.}{Absolute proper motion of IRAS 00259+5625}

\usepackage{times}

\begin{document}
\RelativeFigurePath{/home/sakai/Aips/paper_IRAS05168}

\title{Absolute Proper Motion of IRAS 00259+5625 with VERAF\\ Indication of Superbubble Expansion Motion}
\author{Nobuyuki Sakai$^{1}$, Mayumi Sato$^{2}$, Kazuhito Motogi$^{3}$, Takumi Nagayama$^{4}$, Katsunori M. Shibata$^{1,4}$, Masahiro Kanaguchi$^{4}$, and Mareki Honma$^{1,4}$%
\thanks{Last update: January 19, 2007}}
\affil{%
   $^{1}$The Graduate University for Advanced Studies (Sokendai), Mitaka, Tokyo 181-8588}
\affil{%
$^{2}$Max-Planck-Institut f\"{u}r Radioastronomie, Auf dem H\"{u}gel 69, 53121 Bonn, Germany}
\affil{%
$^{3}$The Research Institute for Time Studies, Yamaguchi University, Yoshida 1677-1, Yamaguchi, Yamaguchi, 753-8511, Japan}
\affil{%
$^{4}$Mizusawa VLBI Observatory, National Astronomical Observatory of Japan, Mitaka, Tokyo 181-8588}

\email{nobuyuki.sakai@nao.ac.jp}
\KeyWords{{\bf Galaxy}:kinematics and dynamics --- {\bf ISM}:superbubbles --- {\bf ISM}:individual (IRAS 00259+5625, LBN 594, CB 3) --- \\{\bf techniques}:interferometric --- {\bf VERA}}

\maketitle

\begin{abstract}
 We present the first measurement of the absolute proper motions of IRAS 00259+5625 (CB3, LBN594) associated with the H{\scriptsize I} loop called the ``NGC281 superbubble" that extends from the Galactic plane over $\sim$ 300 pc toward decreasing galactic latitude. The proper motion components measured with VERA are ($\mu_{\alpha} \mathrm{cos}\delta$,\ $\mu_{\delta}$) = ($-$2.48 $\pm$ 0.32,\ $-$2.85 $\pm$ 0.65) mas yr$^{-1}$, converted into ($\mu_{l} \mathrm{cos}b$,\ $\mu_{b}$) = ($-$2.72 $\pm$ 0.32,\ $-$2.62 $\pm$ 0.65) mas yr$^{-1}$ in the Galactic coordinates. The measured proper motion perpendicular to the Galactic plane ($\mu_{b}$) shows vertical motion away from the Galactic plane with a significance of about $\sim$4-$\sigma$. As for the source distance, the distance measured with VERA is marginal, 2.4$^{+1.0}_{-0.6}$ kpc. Using the distance, an absolute vertical motion ($v_{b}$) of $-$17.9 $\pm$ 12.2 km s$^{-1}$ is determined with $\sim$1.5-$\sigma$ significance. The tendency of the large vertical motion is consistent with previous VLBI results for NGC 281 associated with the same superbubble. Thus, our VLBI results indicate the superbubble expansion motion whose origin is believed to be sequential supernova explosions. 
\end{abstract}

\section{Introduction}
 Since their first discovery by Heiles (1979), shell or arc-like H{\scriptsize I} objects with sizes ranging from a few pc to more than 1 kpc have been discovered in the Milky Way Galaxy. Such objects have also been identified in many external spiral or irregular galaxies including the LMC and SMC (e.g., Brinks $\&$ Bajaja 1986; Deul $\&$ den Hartog 1990; Kim et al. 1998; Staveley-Smith et al. 1997; Bagetakos et al. 2011). In particular, the large H{\scriptsize I} objects (up to a few kiloparsecs in scale) are called superbubbles or supershells. Two main models are proposed for the origin of the superbubbles/supershells: (1) stellar winds and sequential supernova explosions in an OB association (e.g., Tenorio-Tagle $\&$ Bodenheimer 1988) and (2) collision of a high-velocity cloud with the Galactic disk (e.g., Tenorio-Tagle 1991). Thus, superbubbles/supershells are believed to play a role in mass and energy transportation from the disk to the halo and are called galactic ``fountains'' (Shapiro $\&$ Field 1976) or ``chimneys'' (Norman $\&$ Ikeuchi, 1989). 

 In the Galaxy, a small number of chimneys and fragmenting superbubbles/supershells (on several hundred to one thousand parsec scales) have been studied: e.g., the Orion-Eridanus superbubble (Cowie, et al. 1979), the Cygnus superbubble (Cash, et al. 1980), the Stockert chimney (M\"{\rm{u}}ller, et al. 1987), the Aquila supershell (Maciejewski et al. 1996), the Scutum supershell (Callaway, et al. 2000), the W4 chimney/superbubble (Normandeau et al. 1996), the Ophiuchus superbubble (Pidopryhora et al. 2007), and the NGC281 superbubble (Megeath et al. 2002, 2003). However, these studies were mainly conducted using 2D positions on the sky and 1D line-of-sight velocities of the objects, meaning that there are uncertainties in terms of the origin and precise physical parameters (e.g., size and expansion velocity) of the superbubbles.

\begin{table*}[th]
\begin{center}
\small
\caption{Observations}
\begin{tabular}{lcccc}
\hline
\hline
Epoch	&Date	&Time Range	&Beam	&Detected maser feature\\ 
  		&	&(UTC)		&(mas)	&	\\
\hline
  A	&2008 Jan 04&11:35-20:45 &1.19$\times$0.83 @127$\fdg$8  &1, 2a-b, 6, 7a\\
  B	&2008 Feb 22&01:00-11:00 &1.10$\times$0.74 @129$\fdg$5  &1, 2a-b, 6, 7a\\
  C	&2008 Mar 23&21:00-07:45 &1.26$\times$0.97 @182$\fdg$6  &1, 2a-b\\
  D	&2008 Apr 29&19:00-05:00 &1.33$\times$0.87 @150$\fdg$9  &1, 2a-b\\
  E	&2008 Jul 06&15:00-01:40 &1.16$\times$0.76 @147$\fdg$0  &1, 2a\\
  F	&2008 Aug 02&12:00-22:40 &1.22$\times$0.83 @156$\fdg$6  &1, 2a\\
  G	&2008 Sep 08&11:00-21:40 &1.34$\times$0.70 @182$\fdg$9  &\\
  H	&2008 Nov 23&06:00-15:10 &1.12$\times$0.73 @141$\fdg$7  &4a-b, 5\\
  I	&2008 Dec 26&04:00-13:00 &1.17$\times$0.84 @148$\fdg$7  &3, 4a-b, 5\\
  J	&2009 Feb 10&01:00-10:10 &1.21$\times$0.78 @139$\fdg$0  &3, 4a-c, 5, 7b\\
  K	&2009 Mar 08&23:00-08:10 &1.22$\times$0.73 @133$\fdg$9  &3, 4a, 4c, 7b\\
  L	&2009 May 07&19:35-04:45 &1.20$\times$0.79 @152$\fdg$5  &3, 4c\\
  M	&2009 Sep 05&11:35-20:45 &1.40$\times$0.68 @145$\fdg$7  &\\

\hline
\multicolumn{0}{@{}l@{}}{\hbox to 0pt{\parbox{180mm}{\footnotesize
\par\noindent
}\hss}}
\end{tabular}
\end{center}
\end{table*}

\begin{table*}[th]
%\begin{center}
\caption{Source Data}
\begin{tabular}{lccccc}
\hline
\hline
Source Name	&R.A.	&Decl.	&S.A.\footnotemark[$*$]	&Flux Density	&Note	\\
		&(J2000.0)	&(J2000.0)				&($^\circ$)			&(Jy)	&\\
\hline
IRAS 00259+5625	&\timeform{00h28m43.5075s}\footnotemark[$\dag$]	&\timeform{56d41'56".868}\footnotemark[$\dag$]	&	&0.8 $\sim$49.3	&H$_{2}$O masers	\\
J0042+5708		&\quad \timeform{00h42m19.4517s}\footnotemark[$\ddag$]        &\timeform{57d08'36".586}\footnotemark[$\ddag$]	&1.908	&0.12$\sim$0.33	&Phase-reference calibrator	\\
\hline
\multicolumn{4}{@{}l@{}}{\hbox to 0pt{\parbox{180mm}{\footnotesize
\par\noindent
\footnotemark[$*$]The separation angle between the maser and the reference sources.\\
\footnotemark[$\dag$]The tracking positions are shifted $\sim$9$\timeform{"}$ with respect to the maser-detected position of feature 2b at epoch A, ($\alpha$, $\delta$)=(\timeform{00h28m42s.5998}, \timeform{56d42'01".098}, J2000).\\
\footnotemark[$\ddag$]The positions are based on Beasley et al. (2002).
}\hss}}
\end{tabular}
%\end{center}
\end{table*}

 Recently, the first VLBI studies were conducted toward one of the superbubbles, the ``NGC281 superbubble'' located $\sim$ 300 pc off the midplane of the Perseus arm, to determine 3D positions and motions of star-forming regions associated with the superbubble (Sato et al. 2007, 2008; Moellenbrock et al. 2009). Sato et al. (2007) showed the direct evidence of the superbubble expansion motion based on absolute proper motions of NGC 281 measured with VERA (VLBI Exploration of Radio Astrometry). The reported proper motions were ($\mu_{l} \mathrm{cos}b$,\ $\mu_{b}$) = ($-$2.88 $\pm$ 0.18,\ $-$2.66 $\pm$ 0.26) mas yr$^{-1}$ in the Galactic coordinates, which clearly shows the vertical motion with $\sim$ 10-$\sigma$ significance. Following this, Moellenbrock et al. (2009) used the VLBA (Very Long Baseline Array) to measure proper motions and parallactic distance of IRAS 00420+5530, a star-forming region associated with the NGC281 superbubble. Also, Sato et al. (2008) reported updated proper motions and parallactic distance of NGC 281 with observations additional to those of Sato et al. (2007), and they discussed 3D structure and kinematics of the NGC281 superbubble based on the VLBI results (see details in Sato et al. 2008). Thus, VLBI study is crucial for understanding not only the origin of the superbubbles, but also interaction between the disk and the halo, which may be related to the Galaxy evolution. To understand the superbubbles with more astrometric data, we conducted VERA observations toward an H$_{2}$O maser emission of IRAS 00259+5625, another star-forming region associated with the NGC281 superbubble.

 IRAS 00259+5625 is reported as an intermediate-mass star-forming region (Codella and Bachiller 1999) and sometimes referred to as LBN 594 and CB3 (Lynds 1965; Clemens and Barvainis 1988). The name of CB3 is based on the catalog of the Bok globules in Clemens and Barvainis (1988). IRAS 00259+5625 has also been investigated with different wavelengths at near-infrared (Yun and Clemens 1995), millimeter (Launhardt and Henning 1997), and submillimeter wavelengths (Launhardt et al. 1997), which showed that there were several sites of star formation with continuum emissions of slightly different positions in the source. In addition, a molecular bipolar outflow elongated from north to south has been detected in the source (Yun and Clemens 1994; Codella and Bachiller 1999). H$_{2}$O masers were also detected in the source with single-dish observations (Scappini et al. 1991) and VLA interferometric observation (de Gregorio-Monsalvo et al. 2006). However, no proper motion measurements with multi-epoch interferometric observations have been reported for the source. Using H$_{2}$O masers in the source, we have conducted a VLBI study to understand kinematics and dynamics of the source related to the superbubble. In the present paper we report on astrometric observations (especially for proper motions) of IRAS 00259+5625 with VERA.

\section{Observations and Data Reduction}
\subsection{VLBI Observations with VERA}
 Between January 2008 and September 2009, we carried out 13 epoch observations of H$_{2}$O maser line at a rest frequency of 22.235080 GHz to measure parallax and proper motions of IRAS 00259+5625. Details of dates for the 13 observations are listed in table 1. 
Typical synthesized beam was 1.2 $\times$ 0.8 mas with a position angle of 149$^{\circ}$.
We also observed reference source J0042+5708 with the target source for the phase-referencing observation. 
Both the target (maser) and the reference (QSO) sources were observed simultaneously using the dual-beam mode (Kobayashi et al. 2008). 
Fringe-finder sources J2238+1242 and CTA 102 were observed and used for calibration of clock parameters in correlation processing of all 13 observations. 
On-source time for IRAS 00259+5625 was $\sim$ 6.5 hours in every observation, compared with a total observation time of $\sim$ 9.4 hours. In table 2 we summarize the tracking centers of the maser and the reference sources with the separation angle between the two sources.

 In these observations, left-handed circular polarization was recorded onto magnetic tapes at a rate of 1024 Mbps with 2-bit quantization after filtering was performed using the VERA digital filter. 
Total bandwidth of 256 MHz consisted of 16 of 16-MHz IF sub-bands. One of the 16 IFs was assigned for the maser source, and the other 15 IFs were assigned for the continuum sources such as the position reference and fringe-finder sources.
Magnetic tapes recorded at the four VERA stations were delivered to NAOJ Mitaka to conduct correlation processing with the Mitaka FX correlator. 
The correlator accumulation period was one second. To achieve high-frequency (velocity) resolution for the maser source, we used only a bandwidth of 8 MHz with 512 channels assigned for the source in the correlation processing. This led to a frequency resolution of 15.63 kHz and a velocity resolution of 0.21 km s$^{-1}$ for the maser source. 
In contrast, each of the 15 IFs was composed of 64 channels for the continuum sources.

\subsection{Data Reduction}
 The Astronomical Image Processing System (AIPS, NRAO) was used for data calibration, and general phase-referencing analysis was applied to determine absolute maser positions in the same manner described in Sakai et al. (2012). First, delay model correction was conducted with a precise geodetic model, the most updated Earth-rotation parameters provided by IERS, tropospheric delays measured with GPS receivers at each VERA station (Honma et al. 2008a), and ionospheric delays based on the Global Ionosphere Map (GIM), which were produced every two hours by the University of Bern. Second, fringe search was conducted with the fringe finders J2238+1242 and CTA 102 to determine the clock offsets between each VERA station. Third, a second fringe search and self-calibration imaging were conducted with the position reference J0042+5708 by referring to the clock offsets determined with J2238+1242 and CTA 102. Fourth, we transferred the complex gain solved for the position reference to the maser source IRAS 00259+5625 through the correction of the instrumental phase difference between the dual-beam system with VERA (Honma et al. 2008b). Finally, Fourier transform of the corrected visibilities and deconvolution were conducted to create both dirty and CLEANed images of each maser channel using the ``imagr" task in AIPS. After the data calibration, we used the ``jmfit" task in AIPS to determine an absolute maser position and a flux of the maser source by elliptical Gaussians fitting to the brightness peak of the CLEANed map. 

 We regarded a maser as detected if it achieved a high SNR (signal-to-noise ratio) of five or more. To identify the same maser spot in each observation epoch for astrometry, we selected the maser spot detected in the same velocity channel with continuous observation epochs. Additionally, to select the same maser spot from a multiple masers map, we checked the position differences of a selected maser with a proper motion threshold of 15 mas yr$^{-1}$ during observations: the position differences should not exceed the threshold. Note that the threshold was chosen roughly considering Galactic and assumed maser internal motions, meaning that the Galactic rotation model and line-of-sight velocity range may be used for the order estimation of the threshold. Among the maser spots identified using the criteria, spots detected in three epochs or more were used for proper motion determinations,  while spots detected in four epochs or more with two or more continuous velocity channels were used for parallax determinations. 

 As a result of data analyses with the above criteria, we detected three maser features to determine parallaxes in table 3; six maser features were detected to determine proper motions in table 4. Note that the feature is recognized as a cluster composed of continuous velocity channels. In a model fitting process for the parallax and the proper motion determinations, we used ``VERA$\underline{\hspace{0.5em}}$Parallax", one of the tasks performed by the ``VEDA (VEra Data Analyzer)", data analyzing software developed at NAOJ. In the task, we assumed that source motions can be described by a combination of linear proper and sinusoidal parallax motions.

\begin{figure}[htbp]
  \begin{center}
      \includegraphics[width=180mm, height=197mm]{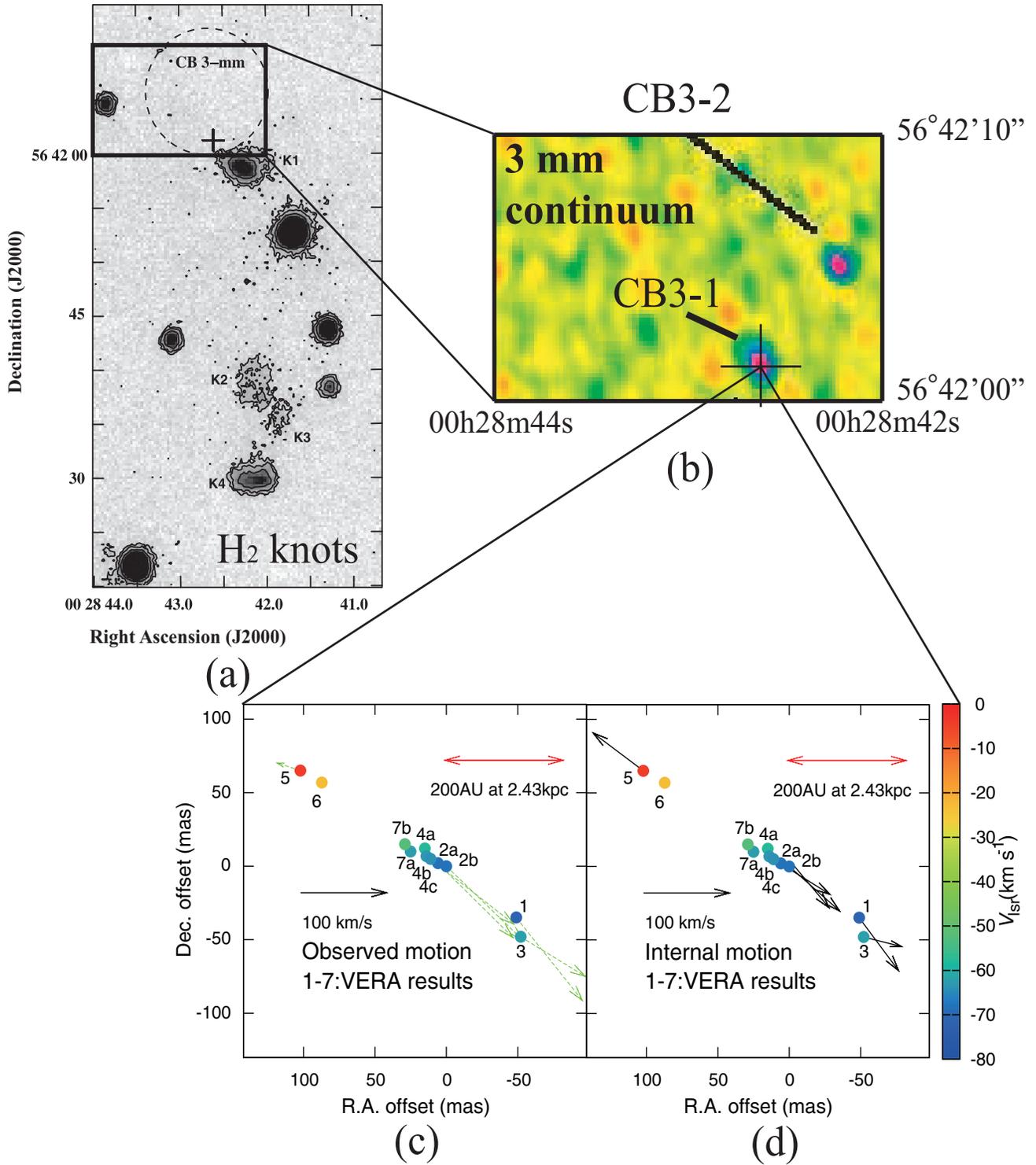}
\end{center}
\caption{\vtop{\hsize=450pt (a) Large-scale view of IRAS 00259+5625 (CB3) with H$_{2}$ knots as K1, K2, K3, and K4 and millimeter source as CB3-mm (Massi et al. 2004; Launhardt and Henning 1997). The positions of K1 and K2 knots are almost consistent with those of integrated SiO (J=5-4) emission peaks at $-$42.5 km s$^{-1}$ regarded as a blue-shifted component of molecular bi-polar outflow (Massi et al. 2004). Dashed circle represents the beam size of the millimeter observations ($\simeq$12$\timeform{"}$, Launhardt and Henning 1997). Cross represents the maser-detected position in table 2. (b) Magnified view of the rectangle in fig. 1a with two millimeter (3 mm) sources as CB3-1 and CB3-2 (Fuente et al. 2007). Cross is the maser-detected position described above. (c) Maser distribution map. The nominal origin is set to the maser-detected position in table 2. Green vectors represent direct observed motions with respect to the position reference source in table 2. (d) Same as (c), but with the internal motions. The black vectors show the internal motions with the systematic motions subtracted (see text).}}
\label{fig1}
\end{figure}

\clearpage

\begin{figure}[htbp]
  \begin{center}
      \includegraphics[width=180mm, height=206mm]{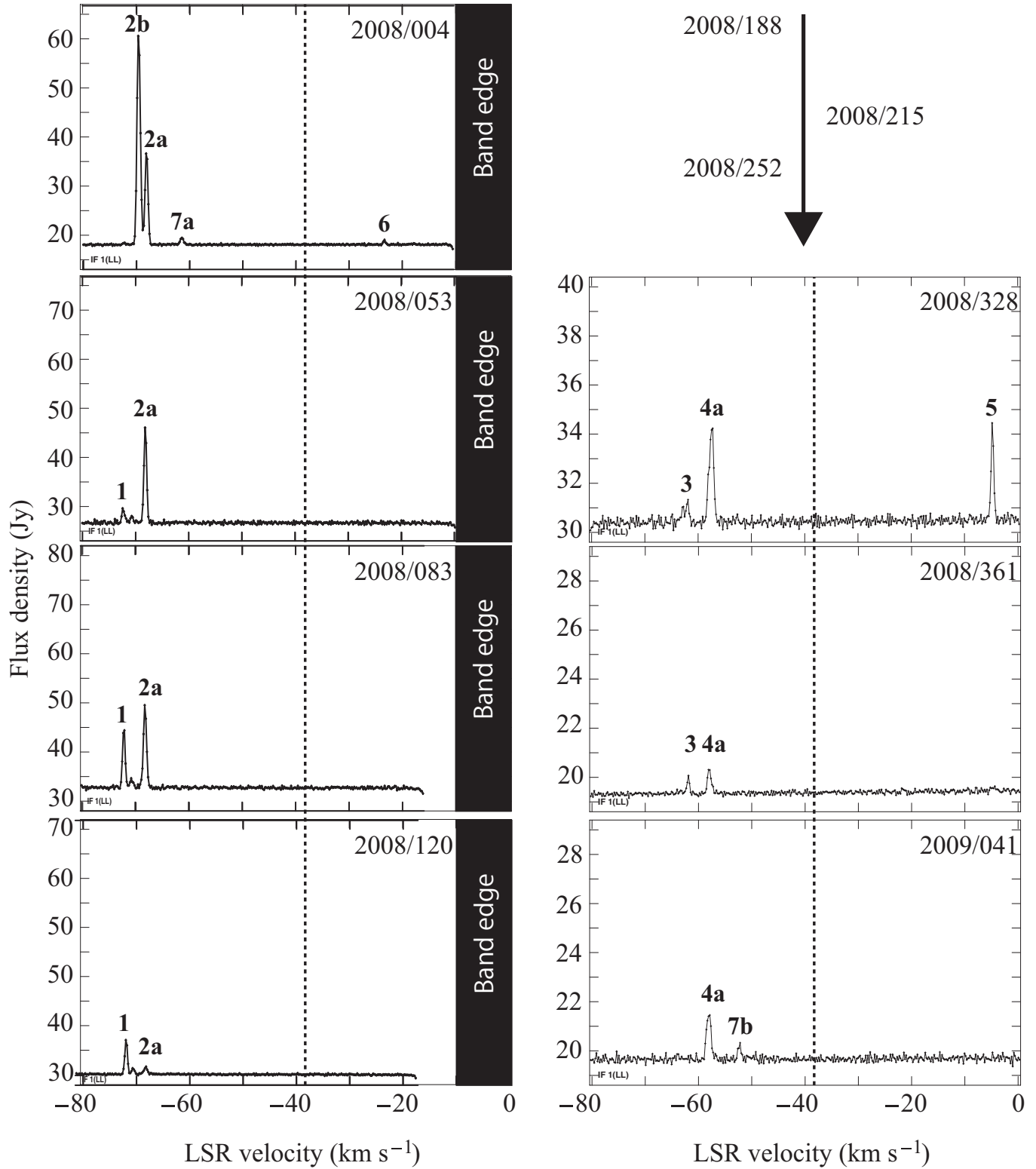}
\end{center}
\caption{\vtop{\hsize=450pt Time variation of scalar averaged cross power spectrum for IRAS 00259+5625 with several baselines averaged. Observation dates (yyyy/ddd) are plotted on upper right corners of each panel. Dotted lines in each panel show a systemic velocity of IRAS 00259+5625, $V_{\rm{LSR}}$ = $-$38.3 km s$^{-1}$ of $^{12}$CO (J=2-1) in Clemens and Barvainis (1988). The filled black area shows band edge of a spectrum. Note that velocity coverages differ slightly between each observation.}}
\label{fig2}
\end{figure}

\clearpage

\begin{table*}
\begin{center}
\caption{Parallax fits.\footnotemark[$*$]}
\small
\begin{tabular}{ccccccc}
\hline
\hline
Feature	&$V_{\rm{LSR}}$	&N$_{epochs}$	&Epochs	&Parallax(Error)			&\multicolumn{2}{c}{Errors	}\\
 \cline{6-7}
		&km s$^{-1}$	&				&			&(mas)		&R.A.	&Dec.	\\
		&	&				&			&		&\multicolumn{2}{c}{(mas)}	\\
\hline
1	&$-$70.3	&6	&ABCDEF+++++++		&			&	&		\\
	&$-$70.5	&6	&ABCDEF+++++++		&			&	&		\\
%	&$-$70.8	&5	&ABCDE++++++++		&	-		&      -		&      -		&	&-	&-		\\
	&$-$71.0	&6	&ABCDEF+++++++		&			&	&		\\
	&$-$71.2	&6	&ABCDEF+++++++		&			&	&		\\
%	&$-$71.4	&6	&ABCDEF+++++++		&			&	&		\\
	&$-$71.6	&6	&ABCDEF+++++++		&			&	&		\\
	&$-$71.8	&6	&ABCDEF+++++++		&			&	&		\\
	&$-$72.0	&6	&ABCDEF+++++++		&			&	&		\\
	&$-$72.2	&6	&ABCDEF+++++++		&			&	&		\\
	&$-$72.4	&6	&ABCDEF+++++++		&			&	&		\\
	&$-$72.6	&6	&ABCDEF+++++++		&			&	&		\\

2a	&$-$68.2	&6	&ABCDEF+++++++		&			&	&		\\
	&$-$68.4	&6	&ABCDEF+++++++		&			&	&		\\

%3	&$-$61.7	&4	&+++++++HIJK++			&-			&-			&-			&	&-	&-		\\
3	&$-$61.9	&4	&++++++++IJKL+		&			&	&		\\
	&$-$62.1	&4	&++++++++IJKL+		&			&	&		\\

\hline
\multicolumn{2}{l}{Combined fit for 14 spots}					&	&	&0.380(0.056)		&0.14	&0.23		\\
\multicolumn{2}{l}{Combined fit for three features}				&	&	&0.443(0.119)		&0.13	&0.19		\\
\hline
\multicolumn{3}{l}{Final (mean of the two combined fittings)}		&	&0.412(0.123)		&	\\

\hline

\multicolumn{4}{@{}l@{}}{\hbox to 0pt{\parbox{135mm}{\footnotesize
\par\noindent
\\
\footnotemark[$*$]Combined fits were conducted to the data set of two (see text). Final value is determined by taking the mean of the two. 
Error of the final parallax is estimated by combining in quadrature the scatter of the individual parallaxes around the mean ($\pm$0.032 mas) and the error bar of individual parallax
 ($\pm$0.119 mas).
}\hss}}
\end{tabular}
\end{center}
\end{table*}

\begin{figure}[htbp]
\includegraphics[width=180mm, height=180mm]{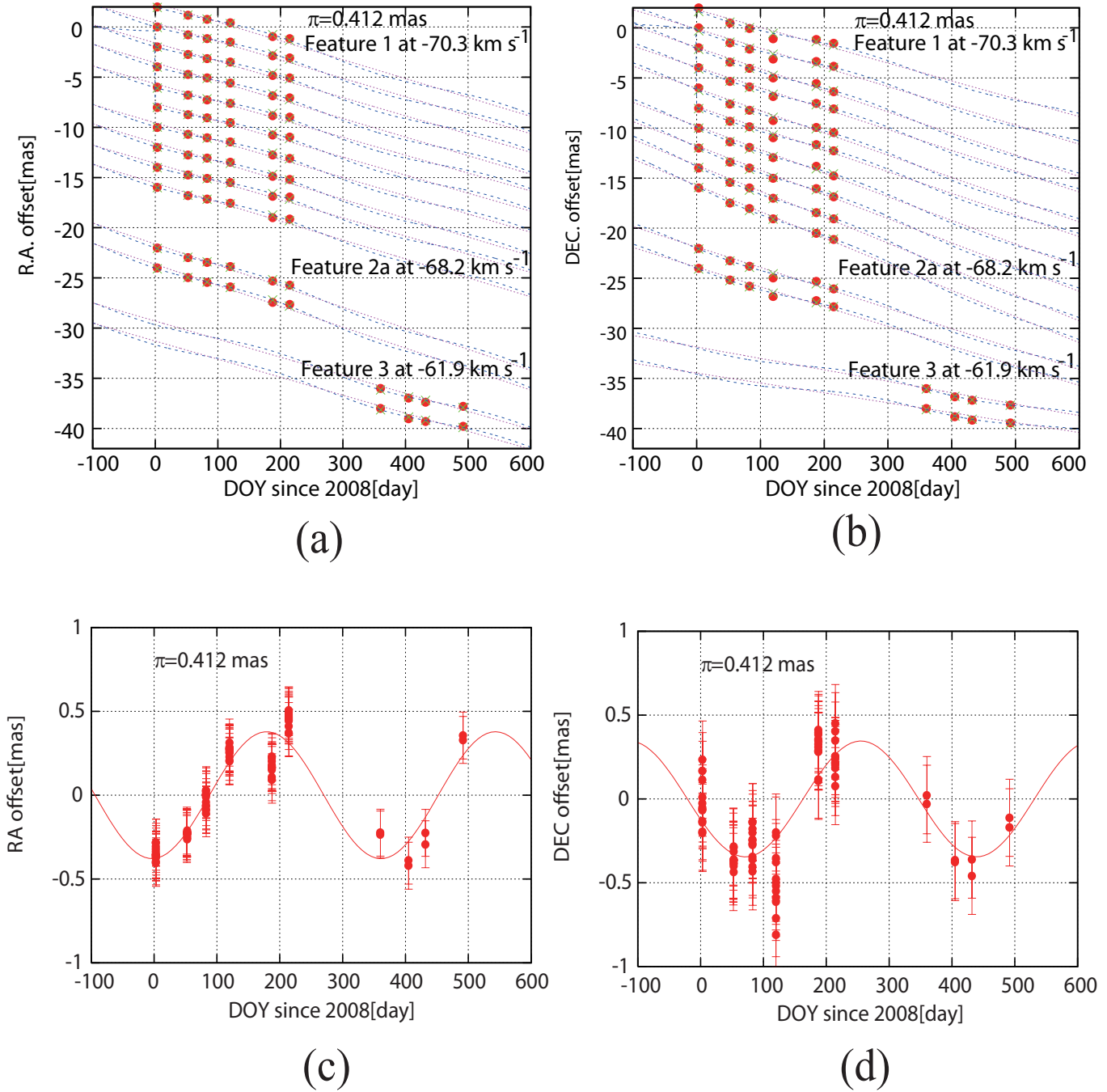}
\caption{\vtop{\hsize=450pt Maser positional evolutions and the combined-fit results for the 14 spots listed in table 3 (see text). The error bars represent position errors resulting from the astrometric (systematic) errors, which are given so that the reduced $\chi^{2}$ becomes unity.
(a) Maser positional evolutions in right ascension with circles. Dotted lines represent proper motions, and dashed lines show fitted lines. Note that offsets for each spot were added arbitrarily to show all spots within the figure. (b) Same as (a), but in declination.
 (c) Parallax motions in right ascension with the proper motions subtracted from (a). The solid curve is fitted to the circles as the parallax motions. (d) Same as (c), but in declination.
}}
\end{figure}

\section{Results}
\subsection{Marginal trigonometric parallax of IRAS 00259+5625}
 During 13 epoch VLBI observations over a period of about 1.6 years, we detected maser emissions in almost all epochs for IRAS 00259+5625 although the time variation of the maser spectrum was significantly large in the observations (e.g., fig. \\ \\ \\ \\ \\ \\ \\ \\ \\ \\ \\ \\ \\ \\ \\ \\ \\ \\ \\ \\ \\ \\ \\ \\ \\ \\ \\ \\ \\ \\ \\ \\ \\ \\ \\ \\ \\ \\ \\ \\ \\ \\ \\ \\ \\ \\ \\ \\ \\ \\ \\ \\ \\
2). The position of the detected maser was consistent with that of the millimeter source (CB3-1) discovered with the Plateau de Bure Interferometer (PdBI) observation in Fuente et al. (2007) as shown in figures 1b and 1c. Figure 1c shows the distribution of 11 detected maser features that were detected in at least one epoch. Among the 11 features, features 1, 2a, and 3 including 14 maser spots were selected for the parallax determination in table 3. However, we could not trace the same maser spot over a year, leading to a relatively large parallax error, although the parallax motions can be clearly seen in the directions of right ascension (R.A.) and declination (Dec.). Using the 14 maser spots, we conducted combined fitting assuming a common parallax with discrete proper motions for each spot. Note that astrometric errors in each epoch are given so that the reduced $\chi^{2}$ becomes unity since a systematic error is generally larger than a thermal error estimate in VLBI observation (e.g., Sanna et al. 2012).

 As a result, the parallax was determined to be 0.380 $\pm$ 0.056 mas for the 14 spots, corresponding to a distance of 2.63$^{+0.45}_{-0.34}$ kpc. However, the 14 spots may not be independent of each other; hence the obtained parallax error (0.056 mas) could be underestimated. To determine a trigonometric parallax of IRAS 00259+5625 with conservative estimation, we again conducted the combined fitting using three bright features (feature 1 with $V_{\rm{LSR}}$ = $-$71.2 km s$^{-1}$, feature 2a with $V_{\rm{LSR}}$ = $-$68.2 km s$^{-1}$, and feature 3 with $V_{\rm{LSR}}$ = $-$61.9 km s$^{-1}$ in table 3). The determined parallax with the representative three features was 0.443 $\pm$ 0.119 mas, corresponding to a distance of 2.26$^{+0.83}_{-0.48}$ kpc. We averaged the two combined fit results to obtain a final parallax result since the 14 spots may not be independent of each other, as we have described above.

 As a consequence, we obtained the final parallax of 0.412 $\pm$ 0.123 mas, corresponding to a distance of 2.43$^{+1.03}_{-0.56}$ kpc. 
Note that the parallax error was estimated by combining in quadrature the scatter of the individual parallaxes around the mean ($\pm$0.032 mas) and the error bar of individual parallax ($\pm$0.119 mas) in the same manner as Honma et al. (2011). As described above, the large error in the final parallax result is mainly due to the short-lived masers (e.g., table 1 and fig. 2). Figures 3a and 3b show the combined fit results (sinusoidal parallax and linear proper motions results) with the final parallax result, fixed for the 14 spots in the directions of R.A. and Dec., respectively.
Error bars in each panel of figure 3 represent the astrometric errors described above. Figures 3c and 3d represent the sinusoidal parallax motions (as circles) with the proper motions subtracted in the directions of R.A. and Dec., respectively. Astrometric errors set for the combined fitting were 0.14 mas for $\Delta$$\alpha$cos$\delta$ and 0.23
mas for $\Delta$$\delta$. These errors mainly originated in the tropospheric zenith delay residuals (e.g., Honma et al. 2007).
Moreover, the separation angle between the maser and the phase reference sources was relatively large (= 1$\fdg$9) in our observations, which caused a large residual of tropospheric zenith delay between the target and the reference pair. 
The error of $\Delta\delta$ is larger than that of $\Delta\alpha$cos$\delta$, consistent with previous VERA results. 

\begin{table*}[htbp]
\begin{center}
\caption{Determination of the systematic proper motions for IRAS 00259+5625.}
\small
\begin{tabular}{lcccc}
\hline
\hline
Feature	&	&\multicolumn{2}{c}{Proper Motions\footnotemark[$\ast$] (Error)}	&Note\\
\cline{3-4}
	&$V_{\rm{LSR}}$	&$\mu_{\alpha} cos\delta$	&$\mu_{\delta}$	&\\
	&km s$^{-1}$	&(mas yr$^{-1}$)&(mas yr$^{-1}$)	&\\
\hline
1 			&$-$70.3 $\sim$ $-$72.6									&$-$6.70	&$-$8.15	&blue-shifted proper motions		\\
2a $\&$ 2b		&$-$68.0 $\sim$ $-$68.4								&$-$7.75	&$-$7.34	&	V	\\
3			&$-$61.9 $\sim$ $-$62.1									&$-$6.57	&$-$3.90	&	V	\\
4c	&$-$63.8 $\sim$ $-$64.2	&$-$8.45	&$-$6.44	&	V	\\
5 		  	&$-$4.0  $\sim$ $-$4.4 									&2.41		&0.76		&red-shifted proper motions		\\
6			&$-$23.1 $\sim$ $-$23.3									&$-$			&$-$			&		\\
7a $\&$ 7b		&$-$61.5 $\sim$ $-$63.0 $\&$ $-$52.1 $\sim$ $-$52.7					&$-$			&$-$			&		\\
4a $\&$ 4b &$-$57.1 $\sim$ $-$58.6 $\&$ $-$63.0 $\sim$ $-$63.4  	&$-$	&$-$	&	\\

\hline
%\multicolumn{2}{l}{Mean-1	(1, 2a$\&$2b, 3, 4a$\&$4b$\&$4c)}							&$-$7.84(0.89)\footnotemark[$\dag$]	&$-$6.70(0.96)\footnotemark[$\dag$]	&blue-shifted proper motions\\

Mean-1 (1, 2a$\&$2b, 3, 4c)&$-$66.5\footnotemark[$\dag$]						&$-$7.37(0.45)\footnotemark[$\ddag$]	&$-$6.46(0.92)\footnotemark[$\ddag$]	&blue-shifted proper motions\\

Mean-2	(Mean-1	$\&$			&$-$35.3\footnotemark[$\dag$]							&$-$2.48(0.32)\footnotemark[$\S$]	&$-$2.85(0.65)\footnotemark[$\S$]	&systamatic proper motions\\
Feature 5) \\

\hline

\multicolumn{4}{@{}l@{}}{\hbox to 0pt{\parbox{170mm}{\footnotesize
\par\noindent \\
\footnotemark[$\ast$]
All proper motions were determined by adapting a parallax of 0.412 mas. All spots included in features 2a and 2b were averaged simultaneously to derive one proper motion pair (see text). \\
\footnotemark[$\dag$]The $V_{\rm{LSR}}$ in Mean-1 was determined by averaging of four representative features as features 1, 2a $\&$ 2b, 3, and 4c, and also the $V_{\rm{LSR}}$ in Mean-2 was determined by averaging of Mean-1 and feature 5 (see text).\\ 
\footnotemark[$\ddag$]The errors were determined by dividing standard deviations by a factor of \ $\sqrt[]{\rm{n}}$, where n is the number of measurements.\\
\footnotemark[$\S$]The errors of the systematic proper motions were determined based on the law of errors propagation. The errors of proper motions for feature 5 were assumed to have the same errors determined by Mean-1.\\
\\
}\hss}}

\end{tabular}
\end{center}
\end{table*}

\subsection{Systematic proper motions of IRAS 00259+5625}
As the next step, we determine the systematic proper motions of IRAS 00259+5625. Note that a maser source has both internal motions (e.g., bi-polar outflow) and systematic motions (e.g., Galactic rotation). Hence, to obtain the systematic motions, one should remove the internal motions. In figure 1c, blue-shifted maser features toward the systemic velocity ($V_{\rm{LSR}}$ = $-$38.3 $\pm$ 3.1 km s$^{-1}$ of $^{\rm{12}}$CO (J=2-1) in Clemens and Barvainis 1988), features 1, 2a-b, 3, 4a-c, and 7a-b, are located around the nominal origin of the map, which was set to be ($\alpha$, $\delta$)=(\timeform{00h28m42s.5998}, \timeform{56d42'01".098}, J2000). Among the 11 maser features in fig. 1c, direct observed proper motions with respect to the position reference were obtained for six features (shown with arrows in fig. 1c). The maser distribution with the velocity vectors indicates that there is a clear sign of bi-polar outflow, which may be connected with molecular bi-polar outflow of IRAS 00259+5625 with a position angle of $\sim$ 0$^{\circ}$ (Yun and Clemens 1994) although position angles differ (fig. 1c). The origin of the molecular bi-polar outflow was proposed to be the millimeter source called CB3-mm (Codella and Bachiller 1999) in fig. 1a. As for the bi-polar outflows at star-forming regions in smaller scales, the bi-polar outflows traced by H$_{2}$O masers have often been seen in many star-forming regions (e.g., Sato, et al. 2010). Thus, here we assume that the internal motion of IRAS 00259+5625 can be modeled by the bi-polar outflow. 

 First, we identified outflow components of red-shifted maser as feature 5 and blue-shifted masers as features 1, 2a-b, 3, and 4c in fig. 1c and table 4. Second, we determined each proper motion by adapting a parallax of 0.412 mas in table 4. Note that the proper motions and LSR velocities of all spots included in features 2a and 2b were averaged simultaneously to derive an averaged proper motion pair ($\mu_{\alpha} \mathrm{cos}\delta$,\ $\mu_{\delta}$) and $V_{\rm{LSR}}$, since features 2a and 2b are located within $\sim$ 6 mas ($\sim$ 144 AU at a distance of 2.4 kpc) with similar motions. The distribution and motion suggest that features 2a and 2b may be associated with the same gas. Third, we averaged only the proper motions and LSR velocities of the blue-shifted masers, yielding averaged blue-shifted proper motions of ($\mu_{\alpha} \mathrm{cos}\delta$,\ $\mu_{\delta}$) = ($-$7.37 $\pm$ 0.45,\ $-$6.46 $\pm$ 0.92) mas yr$^{-1}$ in the equatorial coordinates with an averaged $V_{\rm{LSR}}$ of $-$65.5 km s$^{-1}$ (table 4). Finally, we averaged both the blue-shifted and red-shifted proper motions to determine the systematic proper motions. 

The obtained systematic proper motion components are ($\mu_{\alpha} \mathrm{cos}\delta$,\ $\mu_{\delta}$) = ($-$2.48 $\pm$ 0.32,\ $-$2.85 $\pm$ 0.65) mas yr$^{-1}$ at an averaged maser velocity of $-$35.3 km s$^{-1}$ in table 4. The averaged $V_{\rm{LSR}}$ was calculated by the averaging of the blue-shifted component (as Mean-1 in table 4) and the red-shifted component (as feature 5 in table 4). Note that the errors of the proper motion components were determined based on the law of errors propagation (see table 4). Figure 1d represents maser internal motions (with arrows) with the systematic motions subtracted, which in fact resembles bi-polar outflow. To cross-check the obtained error components of the systematic motions, we calculated the difference between $V_{\rm{LSR}}$ at $^{12}$CO (J=2-1) emission of $-$38.3 km s$^{-1}$ for IRAS 00259+5625 (Clemens and Barvainis 1988) and the averaged $V_{\rm{LSR}}$ of $-$35.3 km s$^{-1}$ with the maser features in table 4. This indicates that the obtained systematic proper motions for IRAS 00259+5625 could include the error, corresponding to the difference. The difference of 3.0 km s$^{-1}$ is converted to $\sim$ 0.3 mas yr$^{-1}$ by adapting a distance of 2.43 kpc for IRAS 00259+5625.

\begin{table*}
\begin{center}
\caption{Peculiar motions of the sources associated with the NGC281 superbubble.\footnotemark[$\ast$]}
\small
\begin{tabular}{lccccccc}
\hline
\hline
Source		&l		&b		&D	 &$U$		&$V$			&$W$				&Ref.\footnotemark[$\ddag$]	\\
			&deg	&deg	&kpc			&km s$^{-1}$	&km s$^{-1}$		&km s$^{-1}$					&	\\
\hline
IRAS 00259+5625	&119.80	&$-$6.03	&2.43$^{+1.02}_{-0.56}$			&12$\pm$8	&$-6\pm$5	&$-18\pm$12				&1	\\
IRAS 00420+5530	&122.02	&$-$7.07	&2.17$^{+0.05}_{-0.05}$			&$26\pm$11\footnotemark[$\dag$]	&$-12\pm$7\footnotemark[$\dag$]	&$1\pm$4\footnotemark[$\dag$]					&2	\\
NGC 281-W	&123.07	&$-$6.31	&2.82$^{+0.26}_{-0.22}$			&$5\pm$4		&$6\pm$4		&$-13\pm$2					&3	\\
%			&		&			&2.38$^{+0.13}_{-0.12}$	&2.4	&0.99		&$9^{+6}_{-6}$		&$3^{+6}_{-6}$		&$-9^{+3}_{-3}$						&4	&CH$_{3}$OH	maser\\
\hline
\textbf{Sun}	&-	&-		&-	&$U_{\odot}$=11.1$\pm$1		&$V_{\odot}$=12.24$\pm$2			&$W_{\odot}$=7.25$\pm$0.5		&4	\\
\hline
\multicolumn{7}{@{}l@{}}{\hbox to 0pt{\parbox{140mm}{\footnotesize
\par\noindent \\
\footnotemark[$\ast$] $\Theta_{0}$=240 km s$^{-1}$ (Reid and Brunthaler 2004), $R_{0}$=8.33 kpc (Gillessen, et al. 2009), and flat rotation model [$\Theta$($R$) = $\Theta_{0}$] are assumed to determine the peculiar motions. Calculation procedure for the errors of the peculiar motions is the same as that used by Johnson and Soderblom (1987), but with the position of the NGP as ($\alpha$ = \timeform{12h51m26s.2817}, $\delta$ = \timeform{27d07'42".013}, J2000) and the third angle of 122$\fdg$932 referred to Reid et al. (2009).\\
\footnotemark[$\dag$]These peculiar motions are based on ($\mu_{\alpha} \mathrm{cos}\delta$,\ $\mu_{\delta}$) = ($-$3.00 $\pm$ 1.26,\ $-$1.40 $\pm$ 0.39) mas yr$^{-1}$ at averaged maser velocity in table 3 of Moellenbrock et al. (2009) (see text). \\
\footnotemark[$\ddag$]References: (1) This paper; (2) Moellenbrock et al. 2009; (3)Sato et al. 2008; (4)Schonrich, Binney, $\&$ Dehnen (2010).}\hss}}

\end{tabular}
\end{center}
\end{table*}%Table 5

\section{Discussion}
\subsection{3D motion of IRAS 00259+5625} %\label{sec:math-symbols}
 Combining the distance and the systematic proper motions for IRAS 00259+5625 with the systemic velocity provides full space motion of the source, which allows us to determine circular and non-circular (peculiar) motions. The calculation procedure is described in Reid et al. 2009: (i) coordinates conversion from the equatorial coordinates into the Galactic coordinates, (ii) velocity conversion from LSR velocity ($V_{\rm{LSR}}$) into heliocentric velocity ($\it{v}_{\rm{helio}}$), and (iii) corrections of peculiar solar motions ($U_{\odot}$, $V_{\odot}$, $W_{\odot}$) and the Galactic constants ($R_{\rm{0}}$, $\Theta_{0}$). Note that we referred to ($U_{\odot}$=11.1$\pm$1, $V_{\odot}$=12.24$\pm$2, $W_{\odot}$=7.25$\pm$0.5) km s$^{-1}$, $R_{\rm{0}}$=8.33 kpc, and $\Theta_{0}$=240 km s$^{-1}$ based on Schonrich, Binney, $\&$ Dehnen (2010), Gillessen et al. (2009), and Reid and Brunthaler (2004), respectively. As a result of the procedure, first we obtain converted proper motions as ($\mu_{l} \mathrm{cos}b$,\ $\mu_{b}$) = ($-$2.72 $\pm$ 0.32,\ $-$2.62 $\pm$ 0.65) mas yr$^{-1}$ in the Galactic coordinates.

 Second, we derive a rotation velocity ($\Theta$) of 234$\pm$5 km s$^{-1}$ through the procedure described above for the circular motion at the source. Note that the error in the rotation velocity was evaluated considering the errors of the parallax, the proper motion components, and the systemic velocity in the same manner as Johnson and Soderblom (1987). Finally, we obtain non-circular motions as ($U$, $V$, $W$) = (12$\pm$8, $-$6$\pm$5, $-$18$\pm$12) km s$^{-1}$ through the same procedure described above with the flat rotation model of $\rm{\Theta}$($R$) = $\Theta_{0}$. The directions of the peculiar motions are toward the Galactic center ($U$), the Galactic rotation ($V$), and the north Galactic pole ($W$) at the source position. The large errors for the non-circular motions mainly originate in the marginal parallax determination. As described in the Introduction, the source is associated with the NGC281 superbubble positioned $\sim$300 pc off the midplane of the Perseus arm (Sato et al. 2008), which indicates that the obtained peculiar motions may originate in both the Perseus arm and the superbubble. In the next section, we will further discuss the origin of the non-circular motions for IRAS 00259+5625.

\subsection{NGC281 superbubble traced by IRAS 00259+5625, IRAS 00420+5530, and NGC 281} %\label{sec:math-symbols} 
 To compare the peculiar motions of IRAS 00259+5625 with previous VLBI results, we listed VLBI results in table 5. The sources listed in table 5 are associated with the same Galactic superbubble, the NGC281 superbubble. Large negative $W$ for IRAS 00259+5625 and NGC 281 is direct evidence of the superbubble expansion motion. As for a $W$ of 1 km s$^{-1}$ in IRAS 00420+5530, it may originate in different 3D positions and motions of the three sources. Note that we referred to proper motions of ($\mu_{\alpha} \mathrm{cos}\delta$,\ $\mu_{\delta}$) = ($-$3.00 $\pm$ 1.26,\ $-$1.40 $\pm$ 0.39) mas yr$^{-1}$ at averaged maser velocity in table 3 of Moellenbrock et al. (2009) for the peculiar motions of IRAS 00420+5530 in table 5. On the other hand, Moellenbrock et al. (2009) reported proper motions of ($\mu_{\alpha} \mathrm{cos}\delta$,\ $\mu_{\delta}$) = ($-$2.52 $\pm$ 0.05,\ $-$0.84 $\pm$ 0.04) mas yr$^{-1}$ at a maser velocity of $-$46.0 km s$^{-1}$ for the systematic proper motions of IRAS 00420+5530. Based on the fact that Brand et al. (2001) reported the systemic velocity of IRAS 00420+5530 as $V_{\rm{LSR}}$ = $-$50.8$\pm$2.7 km s$^{-1}$ of CS(J=3-2), we referred to the previous proper motions to determine the peculiar motions with conservative estimation. However, we emphasize that the listed peculiar motions for IRAS 00420+5530 at the averaged velocity are consistent with peculiar motions at the maser velocity of $-$46.0 km s$^{-1}$ within error.

 Based on table 5, the three sources were superimposed on the Galactic longitude ($l$) and Galactic latitude ($b$) map with $^{12}$CO (J=1-0) color (Dame, Hartmann, and Thaddeus 2001) and H {\scriptsize I} contour (Hartmann et al. 1997) emissions in fig. 4a. Note that the both emissions were integrated over the velocity range of the Perseus arm ($V_{\rm{LSR}} = -60 \ \rm{to} \ -25 \ \rm{km \ s^{-1}}$). Figure 4a also shows peculiar motions of NGC281, IRAS 00420+5530, and IRAS 00259+5625 (with arrows). Clearly, both NGC 281 and IRAS 00259+5625 represent same peculiar motion away from the Galactic plane although IRAS 00420+5530 shows another tendency as parallel peculiar motion in the $l-b$ diagram. The observational result for IRAS 00420+5530 may be explained by three-dimensional configuration of the three sources with the direction of the expansion flow. 

 To model the expansion flow, Sato et al. (2008) made the Galactic longitude ($l$) and $V_{\rm{LSR}}$ map with a ring model as shown in fig. 4b. As for the ring model in Sato et al. (2008), a radially elongated ring model of $\Delta l \sim$ 300 pc, $\Delta r \sim$ 650 pc, and $\Delta z \sim$ 620 pc was proposed with an expansion velocity of 15 km s$^{-1}$ in fig. 4c of Sato  et al. (2008). Note that the proposed absolute lengths were based on VLBI results of NGC281 and IRAS 00420+5530 (see details in Sato et al. 2008). However, if the magnetic field lies along the Perseus arm (e.g., Han et al. 2006), the radially elongated ring model is not consistent with the MHD simulation result in Tomisaka (1998), which showed that the elongation of the superbubble is along the direction of the magnetic field. Our result for IRAS 00259+5625 cannot confirm the validity of the ring model due to the large distance error. To evaluate the proposed ring model precisely, more accurate astrometric observations are required for the NGC281 superbubble. 

 Figures 4a and 4b with VLBI results can be converted into three-dimensional configuration of the three sources as shown in figures 5a and 5b with previous VLBI observations. Figure 5a represents an edge-on diagram sliced with a Galactic longitude of $\sim$ 121 $^{\circ}$ with the assumed ring center ($d_{\rm{ring}}$ = 2.5 kpc assumed in Sato et al. 2008), and figure 5b represents a face-on diagram with the assumed ring center and previous VLBI results. Observed peculiar motion, modeled expansion flow, and residual between the previous two motions are shown in each panel (with different-colored arrows). In other words, the observed peculiar motions for IRAS 00420+5530, IRAS 00259+5625, and NGC 281 in the superbubble region can be explained by the sum of the expansion flow (green arrow) and the residual (blue arrow). As for the origin of the residual (blue arrow), the Perseus Arm's motion is an important candidate since the three sources are associated with the Perseus arm as illustrated in figures 5a and 5b. However, averaged residual components (averaged blue arrow), ($U_{\rm{3\underline{\hspace{3pt}}sources}}$ = 14 $\pm$ 4, $V_{\rm{3\underline{\hspace{3pt}}sources}}$ = $-$2 $\pm$ 2, $W_{\rm{3\underline{\hspace{3pt}}sources}}$ = 1 $\pm$ 5) km s$^{-1}$, are not consistent with averaged non-circular motions of the other nine sources in the Perseus arm, ($U_{\rm{9\underline{\hspace{3pt}}sources}}$ = 9 $\pm$ 2, $V_{\rm{9\underline{\hspace{3pt}}sources}}$ = $-$19 $\pm$ 2, $W_{\rm{9\underline{\hspace{3pt}}sources}}$ = $-$4 $\pm$ 2) km s$^{-1}$ in fig. 5b. Other candidates for the residual are thought to be non-isotropic expansion of the superbubble and motions of the star-forming regions relative to the superbubble. In fact, fig. 5a may support the latter models (in the blue arrows). To model the peculiar motion of the NGC281 superbubble where there are origins of the peculiar motion (e.g., the Perseus arm, local motion in a star-forming region, and the superbubble), more astrometric observations will be required.  \\ \\

As for future expectations, the VERA project aims to observe several hundred H$_{2}$O maser sources located in star-forming regions including superbubbles for any parallax and proper motions determinations within the next decade. This would give us an accurate understanding of the superbubble regions related to the Galaxy evolution in the near future. \\ \\

 \hspace*{30pt} We are grateful to VERA project members for the support they offered during observations. We would like to thank the referee for carefully reading the manuscript. We would also like to thank Ms. Yolande McLean for conducting English proofreading. This work was financially supported by The Graduate University for Advanced Studies (Sokendai), the National Astronomical Observatory of Japan (NAOJ), and the Grant-in-Aid for the Japan Society for the Promotion of Science Fellows (NS).

%\fi

\clearpage

\begin{figure}[htbp]
  \begin{center}
      \includegraphics[width=180mm, height=198.5mm]{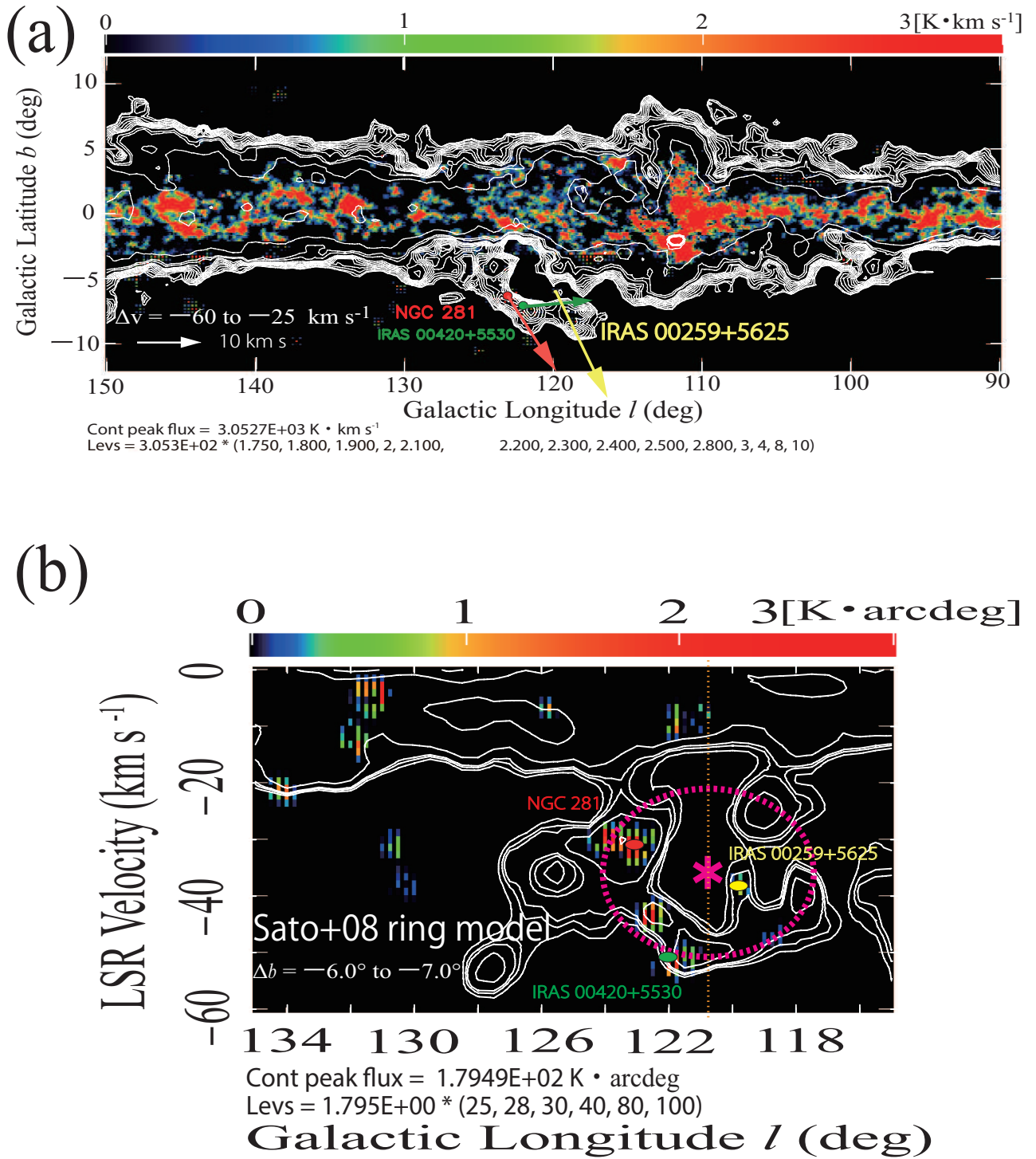}
\end{center}
\caption{\vtop{\hsize=450pt NGC 281 superbubble is shown with white contours H{\tiny I} data from Hartmann and Burton (1997) and color $^{12}$CO (J=1-0) data from Dame, Hartmann, and Thaddeus (2001). The positions of NGC 281 (red), IRAS 00420+5530 (green), and IRAS 00259+5625 (yellow) are indicated. (a) $l$ vs. $b$ map of the region, velocity-integrated for the Perseus-arm line-of-sight velocity range of $V_{\rm{LSR}}$ = $-$60 to $-$25 km s$^{-1}$. The observed non-circular motions of the three sources are plotted as red (NGC 281), green (IRAS 00420+5530), and yellow arrows (IRAS 00259+5625). (b) $l$ vs. $V_{\rm{LSR}}$ diagram, latitude integrated for the galactic latitude range of $b$ = $-$6$^{\circ}$ to $-$7$^{\circ}$. Based on a ring model proposed in Sato et al. (2008), assumed ring center as pink $\ast$ and shape of the ring as pink dotted circle are plotted on the map.}}
\label{fig05}
\end{figure}

\clearpage

\begin{figure}[htbp]
  \begin{center}
      \includegraphics[width=180mm, height=200mm]{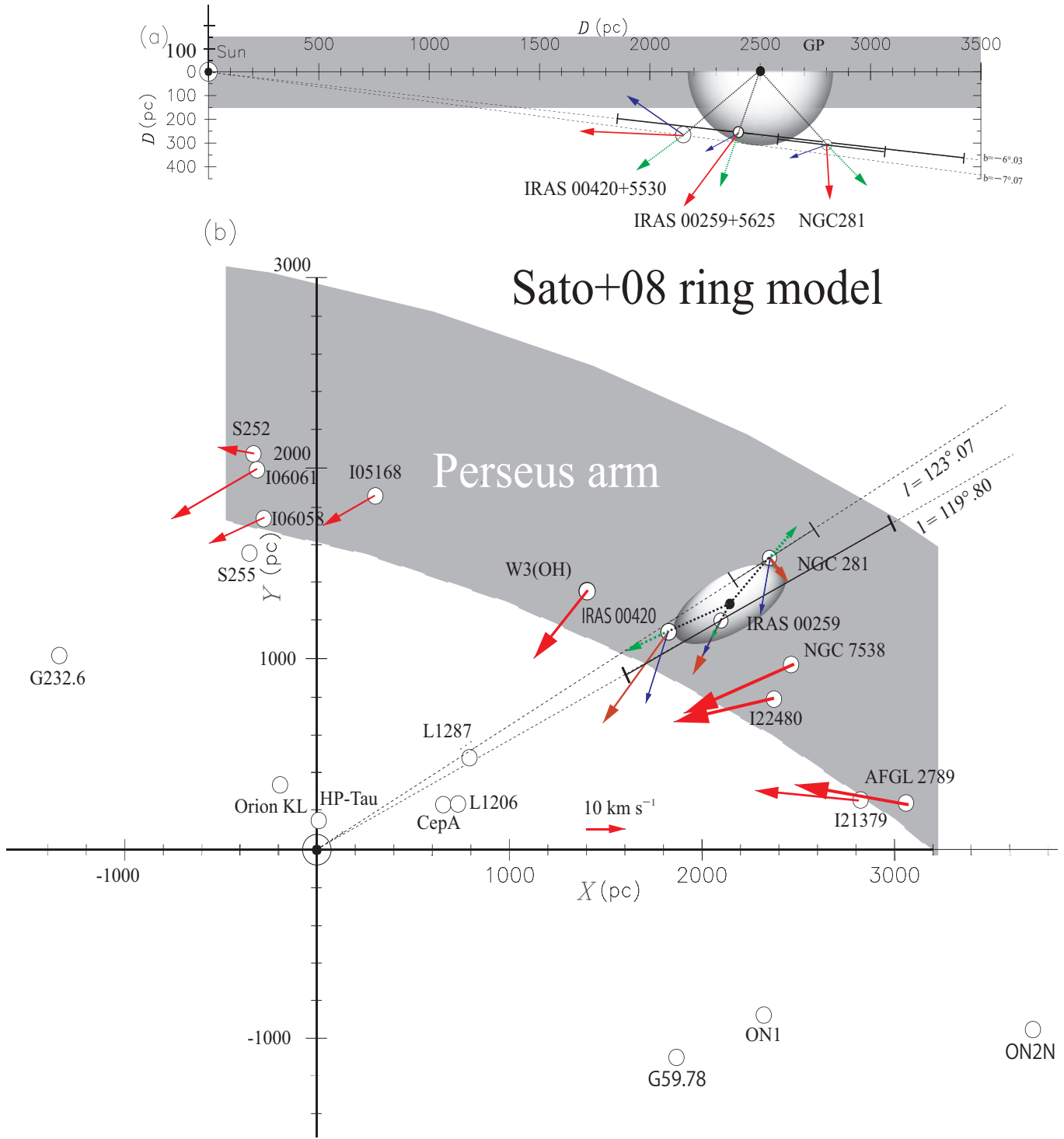}
\end{center}
\caption{\vtop{\hsize=450pt Schematic diagram of the 3D structure and motion of the NGC281 superbubble (Sato, et al. 2008). (a) Edge-on view (at a galactic longitude of $l$ $\sim$ 121$^{\circ}$) and (b) face-on view of the Galactic disk. Note that all arrows represent the non-circular motions with the Galactic rotation and the solar peculiar motions subtracted. Dotted green arrows originate in a ring model proposed in Sato et al. (2008), and blue arrows are residual vectors between the observed and modeled ring motions (see text). Red arrows as the sum of the green and blue arrows show not only observed non-circular motions of the three sources in the NGC 281 superbubble, but also those of nine other sources in the Perseus arm. Note that the nine sources and the other sources with no arrows are previous VLBI results (e.g., summarized in Sakai et al. 2013).}}
\label{fig8}
\end{figure}

\end{document}